\documentclass[prl, twocolumn,amsmath,amssymb]{revtex4}
\usepackage{amssymb}
\usepackage[dvips]{graphicx}
\begin{document}
\title{The bifurcation phenomena in the resistive state of the narrow superconducting channels}

\author{V. V. Baranov$^{1}$, A. G. Balanov$^{2}$, and V. V. Kabanov$^{1}$}

\affiliation{$^{1}$Jozef Stefan Institute, Jamova 39, 1001 Ljubljana,
Slovenia; $^{2}$Department of Physics, Loughborough University, Loughborough, UK}

%\ead{vladimir.baranov@ijs.si, viktor.kabanov@ijs.si}

\begin{abstract}
We have investigated the properties of the resistive state of the narrow superconducting channel of the length $L/\xi=10.88$ on the basis of the time-dependent Ginzburg-Landau model. We have demonstrated that the bifurcation points of the time-dependent Ginzburg-Landau equations cause a number of singularities of the current-voltage characteristic of the channel. We have analytically estimated  the averaged voltage and the period of the oscillating solution for the relatively small currents. We have also found the range of currents where the system possesses the chaotic behavior.
\end{abstract}

\maketitle

It is well-known that the resistive state appears in a superconductor in the range of currents $j_c<j<j_2$. When starting from the superconducting state and increasing the current, it is obvious that the superconducting state switches to the resistive one at a certain critical current. Further increase of the current leads to the switching from the resistive to the normal state at the upper critical current $j_{2}$. In the resistive state the superconductivity and a static electric field coexist in a system. It causes the appearance of phase slip centers (PSCs). The phase of the order parameter (OP) in the resistive state periodically drops by $2\pi$ in a set of sample points. This process leads to the oscillation of the amplitude of the OP. It has been shown that PSCs generate certain singularities on the current-voltage characteristic (CVC). It possesses a stair like structure \cite{iv_kop,sbt}.

The properties of the resistive state of different types of the superconductors have been investigated recently.
 The influence of the ratio between the relaxation times of the time-dependent Ginzburg-Landau (TDGL) \cite{gor_kop} equations on the
 dynamic of the OP in the PSC has been established \cite{mic, vod, vv}. Besides, it has been
 shown that in the limit of strong pair breaking effect due to interaction with phonons the CVC
 demonstrates strong hysteresis in the constant current regime \cite{mic}. All these results
 (see also \cite{brd1, brd2}) demonstrate that different time-periodic and time-quasiperiodic solutions
 arise with the change of the voltage or current \cite{kim}.
Recent experimental observations %%, dmitriev, lucot, tian}
 have revealed the space-time arrangement of the PSCs. Using the low temperature laser scanning microscopy technique it was demonstrated that each voltage jump on the CVC corresponds to generation of a new PSC \cite{sivakov}. The authors have observed the creation of one PSC at a certain critical current. Further increase of the current leads to the spatial rearrangement of the PSCs - two PSCs appear symmetrically with respect to the center of the channel. Finally, the third PSC appears in the middle of the channel (see Fig.2 of Ref.\cite{sivakov}). All these changes of the PSCs' arrangement immediately induce the change of the CVC's slope. However, some properties of the resistive state of a superconductor are not investigated.

 Our paper is structured as follows. First, we introduce the TDGL equations. Then we estimate analytically the averaged voltage and the period of the oscillating solution for the relatively small currents. Finally, we analyze the bifurcation points of the TDGL equations and reveal their properties.

The TDGL model is described by the equations:
\begin{equation}
u\Big(\frac{\partial\psi}{\partial t}+i\phi\psi\Big)=\frac{\partial^2\psi}{\partial x^2}+\psi-\psi |\psi|^2,
\end{equation}
\begin{equation}
j=-\frac{\partial\phi}{\partial x}+\frac{1}{2i}\Big(\psi^*\frac{\partial\psi}{\partial x}-\psi\frac{\partial\psi^*}{\partial x}\Big).
\end{equation}

Here $\psi=\rho\:exp(i\theta)$ is the complex OP, where $\rho$ and $\theta$ are the modulus and phase of the OP respectively. The distance and time are scaled in units of the coherence length
$\xi$ and phase relaxation time $\tau_{\theta}=4\pi\lambda^2\sigma_n/c^2$ respectively, where
$\lambda$ is the penetration depth, $\sigma_n$ is the normal state conductivity, and $c$ is the speed of light. The electrostatic potential $\phi$ is measured in units of $\phi_0/2\pi c\tau_{\theta}$, where $\phi_0=\pi\hbar c/e$ is the flux quantum, $e$ is the electric charge, $\hbar$ is the reduced Planck constant. The current density $j$ is defined
in units of $\phi_0c/8\pi^2\lambda^2\xi$. The only parameter left is $u=\tau_{\rho}/\tau_{\theta}$, where $\tau_{\rho}$ is the relaxation time of the amplitude of the OP. From the other side $u=\xi^2/l_E^2$, where $l_E$ is the penetration depth of the electric field. The penetration depth increases if the inelastic scattering time on phonons is large enough. It can be approximated in the case of gapless superconductivity if we assume that the phenomenological parameter u is small $u<1$ \cite{ivlev_u}.

\begin{figure}
\begin{center}
\includegraphics[angle=-90,width=0.4\textwidth]{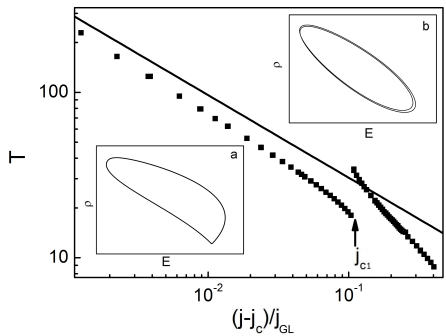}
\end{center}
\caption{The period of the solution as a function of the current. The solid line represents the result of Eq. (3). Arrow indicates the period-doubling
bifurcation point $j_{c1}$. Insets represent projection of the limit cycle trajectory to the $(\rho(0),E(0))$ plane before (a) and after (b) the bifurcation point.
Here $\rho(0)$ is the modulus of the OP, $E(0)$ is the electric field both in the center of the wire.}
\end{figure}

In the present paper we are considering the superconducting wire of a length $L_0/\xi=10.88$ with the following boundary conditions: $\rho(-L/2)=\rho(L/2)=1$ and $d\phi(-L/2)/dxõ=d\phi(L/2)/dx=0$. The absence of the electric field at the end of the wire determines the gradient of the phase: $d\theta(-L/2)/dx=d\theta(L/2)/dx=j$.

Let us start with the features common to the cases of the channels of lengths $L=2L_0, 4L_0$, studied in Ref. \cite{we}. The second TDGL equation in the steady state $j=k(1-k^2)$ possesses two roots when $j<j_c$. One of them is stable and the other is unstable. At $j=j_c$ they collide with each other. It is accompanied by the creation of a limit cycle. Besides, only one Lyapunov exponent crosses zero at this moment \cite{vv}. Therefore, the first singularity of the CVC at $j=j_c$ is a saddle-node homoclinic bifurcation \cite{kuzn}. The period of oscillations in the vicinity of this bifurcation is determined by the formula \cite{we}:
\begin{equation}
T=\frac{\pi\sqrt{3}(u+2)}{2^{3/2}u}\bigl[(j-j_{c})/j_{c}\bigr]^{-1/2}.
\label{period}
\end{equation}

We have solved Eqs.(1) and (2) numerically with the help of the fourth order Runge-Kutta method.  Taking into account the finiteness of the channel, we have obtained that the critical current in our system $j_c$ is not equal to the one in the Ginzburg-Landau theory $j_{GL}=2/3\sqrt{3}$ for an infinite wire. Our calculations show that the critical current $j_c/j_{GL}=1.016$.

The results of our calculations for the period of the oscillating solution together with the analytical estimation (Eq.(3)) are presented in Fig.1. There is a very good agreement of numerical results with Eq.(3) over two orders of magnitude in $(j-j_{c})/j_{c}$.

\begin{figure}
\begin{center}
\includegraphics[angle=-90,width=0.4\textwidth]{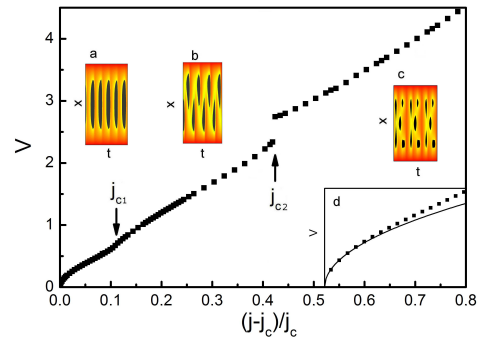}
\end{center}
\caption{The CVC of the channel. Insets a-c show the space-time arrangement of the PSCs in the different regions of the CVC. Inset d represents the CVC for $j<j_{c1}$ in comparison with analytical formula (see the text).}
\end{figure}

From the Josephson relation \cite{iv_kop} and Eq.(3) it follows that the voltage in the vicinity of the saddle-node homoclinic bifurcation is proportional to the square root of the current $V\propto[(j-j_c)/j_{c}]^{1/2}$. Such behavior of the voltage for the region of currents $j_c<j<j_{c1}$ is represented on Fig. 2d. For this range of currents the PSC appears in the middle of the channel (Fig. 2a) in agreement with the experimental results\cite{sivakov}.

Further increase of the current causes the bifurcation of the periodic solution. It causes an increase of the period of the limit cycle at $j=j_{c1}$. On the insets to Fig.1 we have plotted the phase portraits of the system before (inset a) and after (inset b) the bifurcation point in coordinates $(\rho(x=0), E(x=0))$. A single-loop (period-1) limit cycle transforms to a double-loop (period-2) limit cycle at $j_{c1}/j_{GL}=1.125$. As it follows from Fig. 2b, the space-time arrangement of the PSCs is changed: two adjacent PSCs are shifted in the opposite directions with respect to the center of the wire in agreement with the experimental observations \cite{sivakov}. Now the period includes two PSCs. Therefore, the type of the singularity which appears at $j=j_{c1}$ is the period-doubling bifurcation. As a result of the period-doubling, a new frequency $\omega_2 =\omega_1/2$ appears in the spectrum of an electromagnetic radiation generated by the current. It should be noted that the presence of the period-doubling bifurcation in the superconducting channels in the voltage driven regime has been reported earlier \cite{kim}. However, the authors observed the PSCs in the center of the channel, in contrast to our results.

The last, but not the least bifurcation is the destruction of the limit cycle. At $j_{c2}/j_{GL}=1.44$ the limit cycle looses stability. %The space-time arrangement of the PSCs in this area is similar to the current's region $j_{c1}<j<j_{c2}$ (see Fig.2c).
In this case two PSCs situated symmetrically with respect to the center of the wire are accompanied by the third one appearing in the center of the channel (see Fig. 2c). Such arrangement of the PSCs is confirmed by the experimental results \cite{sivakov}. This situation is in contrast to the longest channels cases, where we have obtained the appearance of the third PSC after destruction of the chaos ($L=4L_0$) or inside the range of currents corresponding to the chaotic behavior of the system ($L=2L_0$) \cite{we}. Here we observe the appearance of the chaos and the third PSC simultaneously. We have observed a voltage discontinuity at this bifurcation point. For certain currents from the area $j>j_{c2}$ we have observed so-called "periodic windows", areas of the parameter values, where the limit cycle with a relatively large period becomes stable \cite{anish}. However, it does not induce any singularity on the CVC. For the range of currents $j>j_{c2}$ the solution remains oscillating, but it looses the periodical properties. The Fourier transform of the voltage for $j>j_{c2}$ possesses a lot of different frequencies (see Fig. 3b) instead of $\delta$-function like spectrum (see Fig. 3a) before the singularity. The same behavior can be observed for the cases of the longer channels, studied in Ref. \cite{we}.

\begin{figure}
\begin{center}
\includegraphics[angle=-90,width=0.4\textwidth]{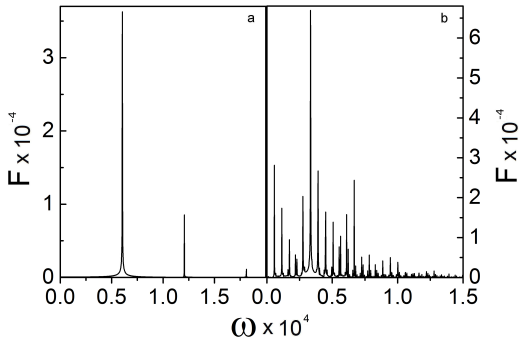}
\end{center}
\caption{The Fourier amplitude of the voltage as a function of the frequency for a) $j<j_{c2}$, b) $j>j_{c2}$. }
\end{figure}

To conclude, we have explored the properties of the narrow superconducting channel. We have demonstrated that the singularities of the CVC  correspond to a number of different bifurcation points of the TDGL equations. The voltage appearance is accompanied by the saddle-node homoclinic bifurcation. It causes the formation of the limit cycle with a diverging period when $j \to j_{c}$.
The voltage $V\propto(j\!-j_c\!)^{1/2}$ in this region. We have also analytically estimated  the period of oscillations in the vicinity of this bifurcation point. The second singularity corresponds to the
period-doubling bifurcation. As a result of this bifurcation, a new frequency equals to the half of the frequency before the bifurcation appears in the spectrum. %We have also proved that the chaos appears via the intermittence.
Finally, the appearance of the third PSC causes the destruction of the limit cycle.

\end{document}